\documentstyle[preprint,aps,epsf]{revtex}
\tightenlines
\begin{document}
\draft
\title{ Classical Many-particle Clusters in Two Dimensions}

\author{G. Date\footnote{email: shyam@imsc.ernet.in},
        M. V. N. Murthy\footnote{email: murthy@imsc.ernet.in} and 
       Radhika Vathsan\footnote{email: radhika@imsc.ernet.in}}

\address
{The Institute of Mathematical Sciences, Madras 600 113, India.\\
}
\date{\today}
\maketitle
\begin{abstract}

We report on a study of a classical, finite system of confined particles
in two dimensions with a two-body repulsive interaction. We first
develop a simple analytical method to obtain equilibrium configurations
 and energies for  few particles.  
When the confinement is harmonic, we prove that the first
transition from a single shell occurs when the number of
particles changes from five to six. The shell structure in the
case of an arbitrary number of particles is shown to be
independent of the strength of the interaction but dependent only
on its functional form. It is also independent of the magnetic
field strength when included. We further study the effect of the
functional form of the confinement potential on the shell
structure. Finally, we report some interesting results when a
three-body interaction is included, albeit in a particular
model.

\end{abstract}

\pacs{PACS numbers: 02.60.Cb, 71.10.-w}

\narrowtext

\section{Introduction}

Two dimensional clusters or ``artificial atoms"  have attracted
considerable attention in recent years \cite{intro}. A cluster of
 a finite number of charged
particles confined by an external potential may be regarded as an
``artificial atom". There are several examples of such systems from
mesoscopic systems to the astroplasma system. In observed systems such as
electrons on a liquid helium film \cite{lhe}, drops of colloidal 
suspensions \cite{csusp} and
confined dusty particles\cite{plasma}, the dynamics is essentially 
classical. On the
other hand, in mesoscopic systems like quantum dots, quantum effects may 
not be negligible\cite{cdot}.  The actual Hamiltonians which capture the 
full dynamics of
these systems are rather complicated. However, several studies assuming
model systems have been carried out.  In particular, there have been
studies on the ordering and transitions of charged particles in
two dimensions\cite{calinon,lozovik,maksym,bedanov,rossler}. 

In a recent Monte Carlo study, Bedanov and Peeters\cite{bedanov} (see also
Bolten and Rossler \cite{rossler}) have analysed the classical ground
state of a system of confined, charged particles interacting through
the Coulomb interaction. By minimising the classical energy, they obtain
numerically the shell structure in a cluster of $N$ particles. They have
systematically listed the shell structure in a ``Mendeleev" table for
$N\le 52$, and for a few large clusters. Similar results are available
also for logarithmic two-body interaction\cite{calinon}. An important fact
that follows from this Mendeleev table is that the charged particles,
confined in a parabolic potential, arrange themselves in concentric shells
where each shell may be thought of as an annulus whose width is much
smaller than the radius. For particular numbers, $N=6,19,37,...$ these
annuli are almost precisely circles. These are called magic numbers. 
There are further systematics (approximate) in this table.  Up to $N=5$,
the particles in the ground state configuration arrange themselves on a
circle.  For $N=6$, there are five particles on the circle and one in the
centre (circle-dot) in the ground state. With the addition of more
particles, a second shell starts forming until the arrangement has 5
particles in the inner shell and 10 particles on the outer shell, ie
$N=15$.  Addition of one more particle ($N=16$), creates a configuration
similar to $N=15$ with the extra particle at the centre. It is interesting 
to note that the multi-shell structure was {\it experimentally} observed by 
Christian Myer\cite{pais} more 
than a hundred years ago. He observed these geometric transitions in a 
system involving magnetic needles floating on water, confined by a bar 
magnet held above the surface. The result, as quoted by J. J. 
Thompson, notes that five magnetic needles always remained on the 
circumference of a circle whereas the sixth one, when added, always 
drifted to the middle. Notice that the interaction between magnetic 
needles is different from Coulomb interaction. Nevertheless the observed 
phenomenon bears resemblance to the numerical results. 

At least up to $N=50$, the geometric transition from 5 to 6, that is
from a circle to a circle-dot, forms a template for the innermost shell.
To a very good approximation, the number of shells for any $N$,
in the Mendeleev table \cite{bedanov}, may be deduced as follows: Write a
given $N$ as a sum of non-repeating multiples of five and a remainder. The
number of terms in the sum is the number of shells. For example, $N=48$
may be written as $N=5+10+15+13$.  This, in our way of looking,
corresponds to having four shells. (It is not expected that this will
hold for large $N$ since the system goes into a hexagonal arrangement in
the bulk.) The main theme of the paper is this geometric transition from
circle to circle-dot configuration when the number of particles changes from 5
to 6. We also derive some general results concerning the shell-structure
for arbitrary $N$. 

In this paper we first consider a cluster of $N$ particles in which the 
repulsive two-body potential is a power-law. 
The classical system we are interested in, to begin with, consists of $N$ 
particles  of equal mass $m$, confined 
in an oscillator potential, in a uniform magnetic field and interacting 
via a two-body potential. The Hamiltonian of such a system of 
particles is given by
\begin{equation}
H = \sum_{i=1}^{N}\left[\frac{(\vec p_i+\vec a_i)^2}{2m} 
+\frac{1}{2}m\omega^2 
r_i^2 \right] + \beta\sum_{i,j (j\ne i)} \frac{1}{(r_{ij}^2)^{\nu}},
\label{ham1}
\end{equation}
where $\vec r_i$ and $\vec p_i$ denote the position and momentum vectors 
of the particle with index $i$. The vector potential, for a uniform magnetic 
field in the symmetric gauge, is given by
\begin{equation}
(a_i)_x = -\omega_L y_i~~; ~~~~~~~~~(a_i)_y =~\omega_L x_i ,
\end{equation}
where $\omega_L$ is the Larmor frequency and $\vec r_{ij} =\vec r_i - 
\vec r_j$. The power $\nu$ (positive) is kept arbitrary. The Coulomb 
Hamiltonian is recovered by choosing $\nu=1/2$.  We also consider the  
case when the two-body interaction is of the form $ -\beta\sum_{i,j(\ne  
i)}\log(r_{ij}^2/\rho^2)$ which is repulsive for $r_{ij}^2 <
\rho^2$ (see Appendix A). This may be closer to the real situation  as  
also the $\nu=1/2$ (Coulomb) case for electrons in a quantum 
dot\cite{zhang}. 
Recently, the $\nu=1$ (inverse square interaction)  case has also been 
analysed in detail\cite{johnson} 
because of its relevance to quantum dot systems. In addition, this case 
nicely lends itself to analytical manipulations.

 We devise a simple
analytical method to obtain the classical ground state energy in {\it two}
steps. First, we extremise the energy for a {\it fixed} total angular
momentum $J$ (which is conserved), and then minimise this energy with
respect to $J$. This has two advantages. It reflects the quantum
degeneracy of the lowest Landau level for electrons in a uniform magnetic
field in the absence of any interaction, at the classical level itself and
secondly it allows one to do the second step of the minimisation over the
quantised values of the angular momentum. Here, 
however, we restrict
ourselves to a classical analysis only and derive some exact results
analytically for the equilibrium configurations. 

In  Section II,  we show that:
\begin{enumerate}
\item{ The configurations 
extremising the energy, modulo overall scale,
are {\it independent} of the parameters of the Hamiltonian 
and also of the angular momentum $J$ so long as the repulsive two-body 
potential falls off as a
power-law with the relative distance and the 
confinement is harmonic. Only the
overall length scale is sensitive to these details. 
}
\item{Two special
configurations, one in which all the $N$ particles are on a circle (referred to
as $\bigcirc$) and the one in which $N - 1$ are on a circle with one at the
center (referred to as $\bigodot$) are always equilibrium
configurations.  We give exact analytical expressions for the
corresponding energy for all $N$ and $\nu$. The $\bigcirc$ has lower energy
for $N \le 5$ while $\bigodot$ has lower energy for $N \ge 6$. In fact the 
$\bigcirc$ is the ground state for all $N\le 5$ while $\bigodot$ is the 
ground state for $N=6,7,8$.  (This is a well known result for Coulomb 
interaction\cite{rossler,bedanov} but is also true in the general case 
considered below for arbitrary $\nu$ up to a maximum value which depends 
on the number of particles.)  
This geometric transition in the ground state is the first one to occur 
and is 
{\it independent} of the precise form of the repulsive interaction. 
}
\item{  
While it is known {\it numerically} that for $N \ge 9$ and for Coulomb
potential\cite{bedanov,rossler} the minimum energy configurations exhibit
approximate multi-shell structure, the special configurations provide an
upper bound on the minimum energy for a whole class of interactions that
we consider here.  The results presented in this section are an
elaboration of our previous work \cite{murthy}.
}
\end{enumerate}

In Section III, we consider the effect of confinement on the geometric 
transition which occurs for $N$ from 5 to 6. The one-body confinement in 
eq.(\ref{ham1}) is 
generalised to $V_{conf} = \sum_{i=1}^N (r_i^2)^{\gamma}$. The oscillator 
confinement is recovered for $\gamma=1$. Surprisingly, we find that there 
exists  a critical value of $\nu$ above which the first geometric 
transition always occurs between 5 and 6 for $\gamma>1$ and between 4 and 
5 for $\gamma<1$. 

In Section IV, we consider the effect of three-body 
perturbations on this geometric transition. We do so by using a 
particular  model 
Hamiltonian whose exact quantum mechanical ground state energy and wave 
function are known in a limit to be defined later. While this again has 
interesting properties in its own right, it is included here 
primarily to study its effect on shell formation. Some 
of questions relating to three-body perturbations are being studied 
and will be published else where\cite{dgm}. 

We conclude in Section V with some 
comments and future prospects. Appendix A contains some general results 
concerning the logarithmic interaction potential. Details of our
 numerical 
simulations are discussed in Appendix B. These are used to check our 
results with ref.\cite{bedanov} for the case of Coulomb potential and 
then extended to other forms of the  potential.  

\section{Geometry of clusters in parabolic confinement}

In this section we consider the classical equilibrium configurations of 
the Hamiltonian given by eq.(\ref{ham1}).  The 
Hamiltonian can be written in terms of dimensionless units by 
introducing a length scale $l= \sqrt{(\hbar/(m\omega)}$ which is the 
basic oscillator length. All distances are measured in terms of this 
basic length unit. Note that  $\hbar$ is introduced only as a 
convenience so that the energy is measured in units of $\hbar\omega$ and 
does not have any other significance as in the quantum case. The analysis 
presented in this paper is entirely classical. The momenta are measured in 
units of $\hbar/l$. 

The new Hamiltonian in these scaled units, but keeping the 
same notation,  may be written as 
\begin{equation}
\frac{H}{\hbar\omega} = \sum_{i=1}^{N}\left[\frac{\vec p_i^{~2}}{2} 
+\frac{1}{2}(1+\alpha^2) r_i^2 + \alpha 
j_i\right] + g\sum_{i,j (j\ne i)} 
\frac{1}{(r_{ij}^2)^{\nu}}, 
\label{ham2}
\end{equation}
 where $j_i = x_ip_{iy}-y_ip_{ix}$, $\alpha = 
\frac{\omega_L}{\omega}$ and $g = \frac{\beta}{\hbar\omega} (l)^{2\nu}$.
Unless otherwise mentioned the summations run from 1 to $N$ hereafter.
While the original coupling constant $\beta$ was dimensional the new 
coupling constant $g$ is dimensionless. Hereafter we assume all the 
energies are measured in units of $\hbar\omega$ and do not write the 
units explicitly. 

To find the equilibrium configurations we carry out the variation in two 
steps. For the first step of the variation we introduce the function
\begin{equation} 
F = H+\lambda(\sum_i j_i - J), \label{eqf}
\end{equation}
where $j_i$ are the single particle 
angular momenta and $\lambda$ is the Lagrange multiplier which enforces
the constraint $J = \sum_i j_i$. 
Setting 
$\delta F=0$, where the variation is done in the full phase space 
variables,  gives the necessary 
equations to determine the equilibrium configuration in the phase space,
\begin{eqnarray}
p_{ix} &=&~(\alpha+\lambda)y_i,   \label{eqpx}\\
p_{iy} &=&-(\alpha+\lambda)x_i,   \label{eqpy}\\  
(1-\lambda^2-2\alpha\lambda)\vec r_i &=&  4g\nu  \sum_{j ( j \ne i)} \frac{\vec 
r_{ij}}{(r_{ij}^2)^{\nu+1}}. \label{eqr}
\end{eqnarray}
This is  the basic set of equations. Any solution to this set of
equations describes an equilibrium configuration which could be
a local minimum/maximum or a saddle point.  We also remark that while the
Hamiltonian is written in a particular symmetric gauge, the variational
equations given above are actually gauge invariant. The main advantage 
of introducing $F$ instead of $H$ for variation is that it allows us to 
keep the dependence of the variational equations on the magnetic field 
even at the classical level. If we vary $H$ separately this advantage is 
lost. We may also include the case of
logarithmic interaction by simply setting the power $\nu=0$
and by setting the prefactor to $4g$ instead of $4g\nu$ in eq.(\ref{eqr}).
To avoid confusion, however, we briefly discuss the results for the
logarithmic case separately in Appendix A.

First we present a qualitative but general analysis of this basic
set of 
equations without making any assumptions. To make the analysis 
simple, we introduce an auxiliary variable, \begin{equation}
\phi = \sum_i r_i^2 . \label{phidef}
\end{equation}
The total angular momentum may now be written in terms of this
auxiliary variable as
\begin{equation}
J = \sum_i (x_i p_{iy}-y_ip_{ix})  = -(\alpha+\lambda)\phi,\label{eqj} 
\end{equation}    
where we have made use of eqs.(\ref{eqpx},\ref{eqpy}).
It is convenient to express $ \vec r_i = R \vec s_i$ where $R$ is a
common scale factor which may be taken to be the radius of the farthest
particle, say the $N^{th}$ one, and $\vec s_i$ denotes the internal
variables.
Therefore 
\begin{equation}
\phi = R^2[\sum_{i=1}^{N-1} s_i^2 +1] \equiv R^2 \widetilde \phi
\end{equation}
since $s_N^2 = 1$ by choice.
Using eq.(\ref{eqr}) and eliminating $\lambda$-dependence using
eq.(\ref{eqj}), we have 
\begin{equation}
\frac{(R^2)^{\nu+1}}{4g\nu}[1+\alpha^2 - \frac{J^2}{R^4 \widetilde
\phi^2}] \vec s_i =  \sum_{j ( j \ne i)} \frac{\vec 
s_{ij}}{(s_{ij}^2)^{\nu+1}}. \label{eqsi}
\end{equation}
Taking the scalar product with $\vec s_i$ and dividing both sides by
$s_i^2 ~(\ne 0)$, we get 
\begin{equation}
\frac{(R^2)^{\nu+1}}{4g\nu}[1+\alpha^2 - \frac{J^2}{R^4 \widetilde
\phi^2}] =  \sum_{j ( j \ne i)} \frac{1 -(s_j/ s_i)
\cos(\theta_{ij})}{(s_i^2+s_j^2-2s_is_j\cos(\theta_{ij}))^{\nu+1}}.
\label{eqdot}  
\end{equation}
Note that the LHS is independent of the particle index $i$. Thus we have $N-1$
independent of equations of the type   
\begin{eqnarray}
\sum_{j ( j \ne i)} \frac{1 -(s_j/ s_i)
\cos(\theta_{ij})}{(s_i^2+s_j^2-2s_is_j\cos(\theta_{ij}))^{\nu+1}}
~~=~~~~~~~~~~~~~~~~~~~~
\nonumber \\
\sum_{j(j\ne k)} \frac{1 -(s_j/ s_k)
\cos(\theta_{kj})}{(s_k^2+s_j^2-2s_ks_j\cos(\theta_{kj}))^{\nu+1}},
~~\forall ~~~k\ne i. \label{eqdt}
\end{eqnarray}
Further by taking the cross product with $\vec s_i$ and dividing by
$s_i$, we get, 
\begin{equation}
\sum_{j ( j \ne i)} \frac{s_j
\sin(\theta_{ij})}{(s_i^2+s_j^2-2s_is_j\cos(\theta_{ij}))^{\nu+1}} = 0,
\label{eqcrs}  
\end{equation}
which provides a further set of $N$ conditions on the internal
coordinates $\vec s_i$. Notice that these conditions are manifestly
scale invariant. Together,
eqs.(\ref{eqdt},\ref{eqcrs}) provide the $2N-1$ necessary
equations for determining the $s_i$ and the angles $\theta_i$. These
 determining 
equations are also completely independent of $\alpha, J$ and $g$. We have,
therefore, the result that $( s_1, s_2,...,s_{N-1}, \theta_1,
\theta_2,...,\theta_{N})$ are independent of the magnetic field
($\alpha$), the 
total angular momentum $J$ and the interaction strength $g$. These
parameters, however, determine the overall scale $R$ through
eq.(\ref{eqdot}).  In fact since $s_N^2 =1$, the corresponding equation
may be taken to be the determining equation for $R$ in terms of the
parameters of the Hamiltonian,
\begin{eqnarray}
\frac{(R^2)^{\nu+1}}{4g\nu}[1+\alpha^2 - \frac{J^2}{R^4 \widetilde
\phi^2}] ~~ = ~~~~~~~~~~~~~~~~~\nonumber \\
\sum_{j=1}^{N-1} \frac{1
-s_j\cos(\theta_{Nj})}{(1+s_j^2-2s_j\cos(\theta_{Nj}))^{\nu+1}} .
 \label{eqscale}  
\end{eqnarray}
Thus we have a {\it very general }
result that the {\it geometry}, that is, the configuration modulo a common 
scale factor,  
or the shell structure of the equilibrium configuration is {\it independent} of
the parameters of the Hamiltonian. The overall size of the system depends 
on the external 
magnetic field strength as well as on the total angular momentum $J$. 
(For further details on the influence of the magnetic field on the 
inter-shell rotation and diffusion, see ref.\cite{schweigert}). The 
shell structure, however, depends on the nature of
the repulsive interaction through the parameter $\nu$ but not its
strength. The above analysis  
is valid even if any one of the $s_i$ is zero, i.e., one particle being at
the origin of the coordinate system, in which case we have two equations
less (not more than one particle can be
at the origin). Note that the statement above applies to {\it all the 
extremum configurations and is  not just restricted to the local minima }. 
In Fig.1 we have shown the typical shell structure for the ground state 
of 25 particles as a function of $J$ and $\nu$. It is easy to see that 
the shell structure, i.e., the number of particles in each shell and 
their 
relative orientation,  is unchanged with $J$ (modulo overall
rotation). The shells get distorted 
with changes in $\nu$. Note that the system size has been normalised to 
the same value in Fig.1 for convenience in display. 

The energy of the equilibrium configuration can be easily computed
by noting that the auxiliary variable $\phi$ defined in
eq.(\ref{phidef}) is related to the two-body potential energy by
\begin{equation}
(1+\alpha^2-\frac{J^2}{\phi^2})\phi=  2g\nu  \sum_{i,j ( j \ne i)} \frac{1}
{(r_{ij}^2)^{\nu}}, \label{eqphi}
\end{equation}
where the RHS is proportional to the potential energy due to
interaction. Since the RHS and $\phi$ are positive definite we have
the condition $(1+\alpha^2)\phi^2 > J^2$. 

\begin{center}
\epsfbox{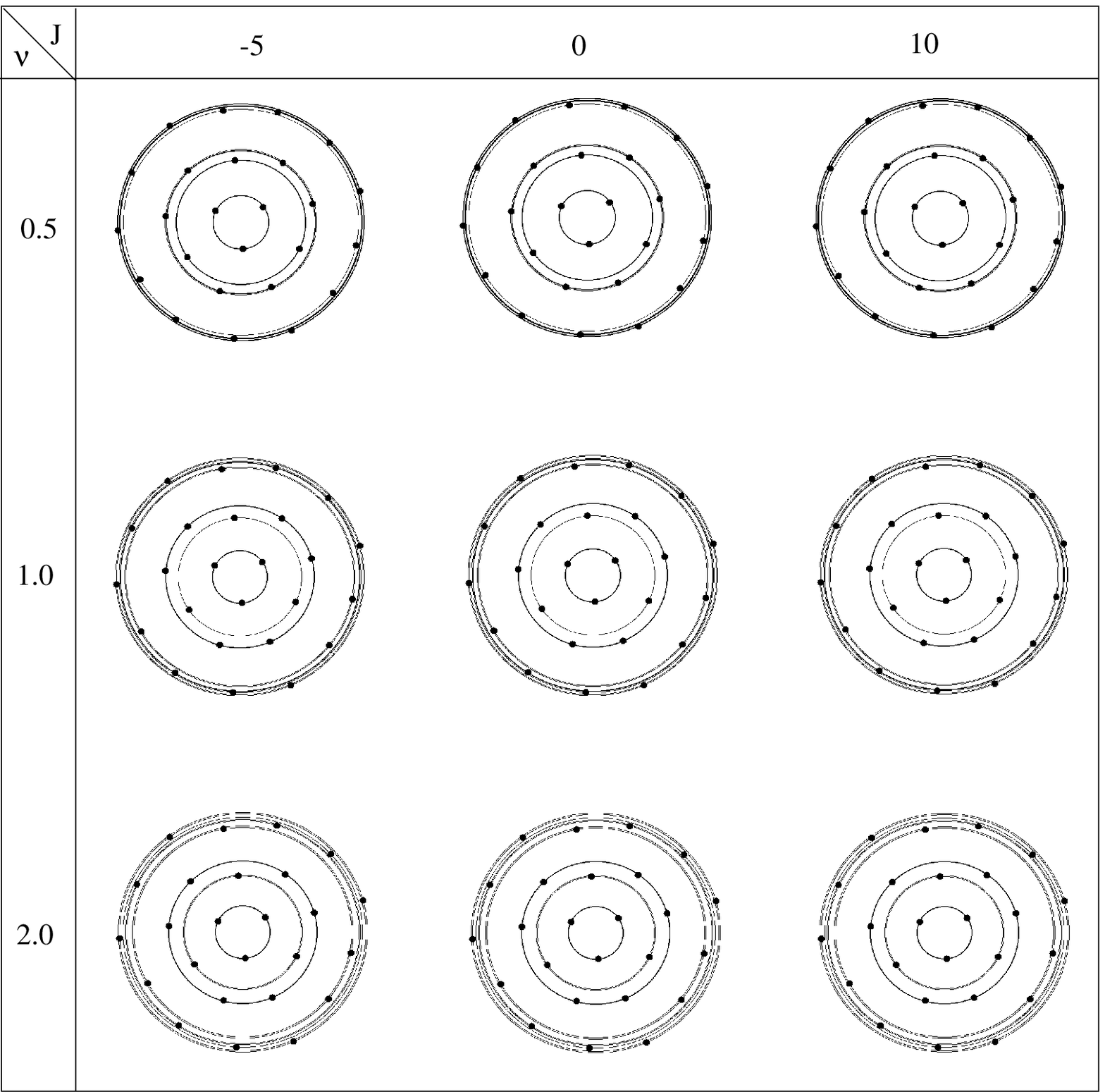}
\begin{tabular}{ll}
{\bf Figure 1:} &
{\sl Shell structure of the ground state for 25
particles as a function of the total}\\
& {\sl angular momentum $J$ and
$\nu$.}
\end{tabular}
\end{center}

In the absence of the magnetic field we may set $J=0$ since 
the true ground state has zero angular momentum. In this limit, the above 
equation reduces to 
\begin{equation}
\frac{1}{2}\sum _i r_i^2 =\nu \left[ g \sum_{i,j ( j \ne i)} \frac{1}
{(r_{ij}^2)^{\nu}} \right],
\end{equation}
which implies that the confinement energy is $\nu$ times the interaction 
energy. For $\nu=1$ both are equal, which is not surprising, since in this 
limit the interaction energy scales like the kinetic energy term. We may 
in fact regard the above statement as a ``virial" theorem valid for 
all equilibrium configurations. 

We have, for the energy at the extrema,
\begin{eqnarray}
E &=& \frac{1}{2}[ (1+\alpha^2)\phi + 2\alpha J + \frac{J^2}{\phi}] +
g\sum_{i,j ( j \ne i)} \frac{1}{(r_{ij}^2)^{\nu}} \nonumber \\
&=&
\frac{\nu+1}{2\nu}(1+\alpha^2)\phi + \alpha J +
\frac{\nu-1}{2\nu}\frac{J^2}{\phi}, \label{eqen}
\end{eqnarray}
where $\phi = R^2\widetilde \phi$ and $\widetilde \phi$ is independent of
$J$. This is the energy of the equilibrium configuration in a
given $J$ sector. A few comments are in order here. Consider the case when 
$g=0$, which corresponds to the case of non-interacting confined 
particles in a magnetic field. This is an exactly solvable problem.  
>From eq.(\ref{eqphi}) we have,
\begin{equation}
(1+\alpha^2-\frac{J^2}{\phi^2})\phi= 0 , 
\end{equation}
which implies either (1) $\phi=0$ and therefore $J =0$ or (2) 
$(1+\alpha^2)\phi^2 = J^2$.  The first of these solutions is the trivial 
solution which one obtains by directly extremising the Hamiltonian. The 
classical solution in this case is independent of the magnetic field. The 
second solution, on the other hand, can be obtained only by imposing the 
condition of fixed $J$ during extremisation. Equivalently, we vary the 
function $F$ and not just $H$. The energy for a given $J$ is then given by
\begin{equation}
E = \sqrt{1+\frac{\omega_L^2}{\omega^2}}~|J| + \frac{\omega_L}{\omega}J,
\end{equation}
which is the energy of confined particles in a magnetic field. Further if 
we remove the confinement potential, the ground state energy is zero for 
all $J\le 0$ which is the solution for the lowest Landau level. (In the 
quantum mechanical case the zero point energy has to be added to the 
solution.). Thus 
the variational method in which the function $F$ is extremised not only 
yields the correct energy but also the correct quantum degeneracy, which 
is infinite in this case.

In the general case when $g \ne 0$, the second step in the extremisation involves extremising  the energy with 
respect to $J$, that is $\partial E/\partial J =0$, 
\begin{eqnarray}
&&\left[ (\nu+1)(1+\alpha^2)R -
(\nu-1)\frac{J^2}{\widetilde\phi^2 R^3}\right]\frac{\partial R}{\partial
J} \nonumber \\
&+& \frac{\nu\alpha}{\widetilde \phi} +
(\nu-1)\frac{J}{\widetilde\phi^2 R^2} =0, 
\end{eqnarray}
since $R$ depends on $J$.
Differentiating $R$ w.r.t. $J$ in eq.(\ref{eqscale}) and cancelling 
the overall powers of $R$, we have
\begin{equation}
\left[ (\nu+1)(1+\alpha^2)R -
(\nu-1)\frac{J^2}{\widetilde\phi^2 R^3}\right]\frac{\partial R}{\partial
J}= \frac{J}{\widetilde\phi^2 R^2}. 
\end{equation}
Eliminating the derivative term in the above equations, we get the 
following solutions at equilibrium for each configuration:
\begin{equation}
J = -\alpha \widetilde \phi R^2; ~~~E  = \frac{\nu+1}{2\nu}\widetilde \phi R^2.
\end{equation}
Substituting for $J$ in eq.(\ref{eqscale}), we get 
\begin{equation}
R^2 = [ 4 g \nu A(\nu,N)]^{\frac{1}{\nu+1}} ,
\end{equation}
where
\begin{equation}
A(\nu,N) \equiv \sum_{j=1}^{N-1} \frac{1
-s_j\cos(\theta_{Nj})}{(1+s_j^2-2s_j\cos(\theta_{Nj}))^{\nu+1}}. 
\end{equation}
An important point to note here is that this energy $E$ is 
{\it independent}
of the magnetic field and its dependence on $g$ is explicit. The
dependence on $N$ and $\nu$ is, however, involved. The angular momentum
$J$ extremising the energy depends on the magnetic field and is zero in
the absence of the magnetic field as it should be. The 
expressions given thus far, though, are valid independent  of the
geometry of the clusters and are exact (for approximate solutions
see eqs.(8,9) in ref.\cite{maksym} for the special case of Coulomb
interaction).   

We now specialise the general results given above to specific
configurations which also happen to be ground states (global minima) for
some $N$. While the results that we obtain are completely analytical and
exact, the choice of configurations is based on the earlier
numerical work \cite{bedanov}. We have also checked the veracity of these
results independently using numerical methods as outlined in 
Appendix B. 

The geometry of the clusters (or shells) is dependent on $\widetilde \phi$
and $A(\nu,N)$, which are as yet unspecified. The equations for the
equilibrium configurations admit many solutions for a given $N$ and $\nu$. 
In particular, there are two special configurations which are always
solutions, viz (i) all the $N$ particles are on a circle, $\bigcirc$ and
(ii) $N-1$ particles are on the circle with one particle at the center,
$\bigodot$. For these two cases, only the overall scale factor $R$ is to
be determined. The angles $\theta_{ij}/2$ are simply multiples of $\pi/N$
and $\pi/(N -1)$ respectively. These configurations, however, need not be
local minima in general. In particular, it has been numerically proved
that for $N\le 5$ the circle configuration is indeed the global minimum
energy configuration (ground state) whereas for $6\le N\le 8$ it is the
circle-dot which is the ground state in the case of Coulomb interaction (
$\nu=1/2$ ). The ground state exhibits multiple shell formation for $N\ge
9$. In what follows we prove analytically that the first transition which
occurs for $N$ from 5 to 6 is independent of $\nu$. However, both
$\bigcirc$ and $\bigodot$ remain ground states for $N=5$ and $N=6$
respectively only for $\nu$ up to a $\nu_{min}$.  In Fig.2, we display the
eigenvalues of the Hessian (matrix of second derivatives) of the
effective potential (see eq(\ref{Veff}))
as a function of
$\nu$.  When $N=5$, the eigenvalues are positive definite up to $\nu=2.4$
for both $\bigcirc$ and $\bigodot$ configurations. Both these
configurations therefore correspond to local minima, but the calculation
of the energy shows clearly that the $\bigcirc$ has lower energy and is
in fact the ground state, confirming the earlier numerical calculations with
$\nu=1/2$. When $N=6$, the eigenvalues are positive all the way up to
$\nu=10$ (modulo a zero eigenvalue related to the rotational invariance)
for the $\bigodot$ configuration. Hence it is a local minimum. However the
$\bigcirc$ configuration ceases to be a local minimum for $\nu>0.484$ and
becomes a saddle point since one of the eigenvalues becomes negative. It 
is possible that there are other configurations, other than the ones 
considered here, which are also local minima. They are not relevant to 
the analysis that is to follow. 

\begin{center}
\epsfbox{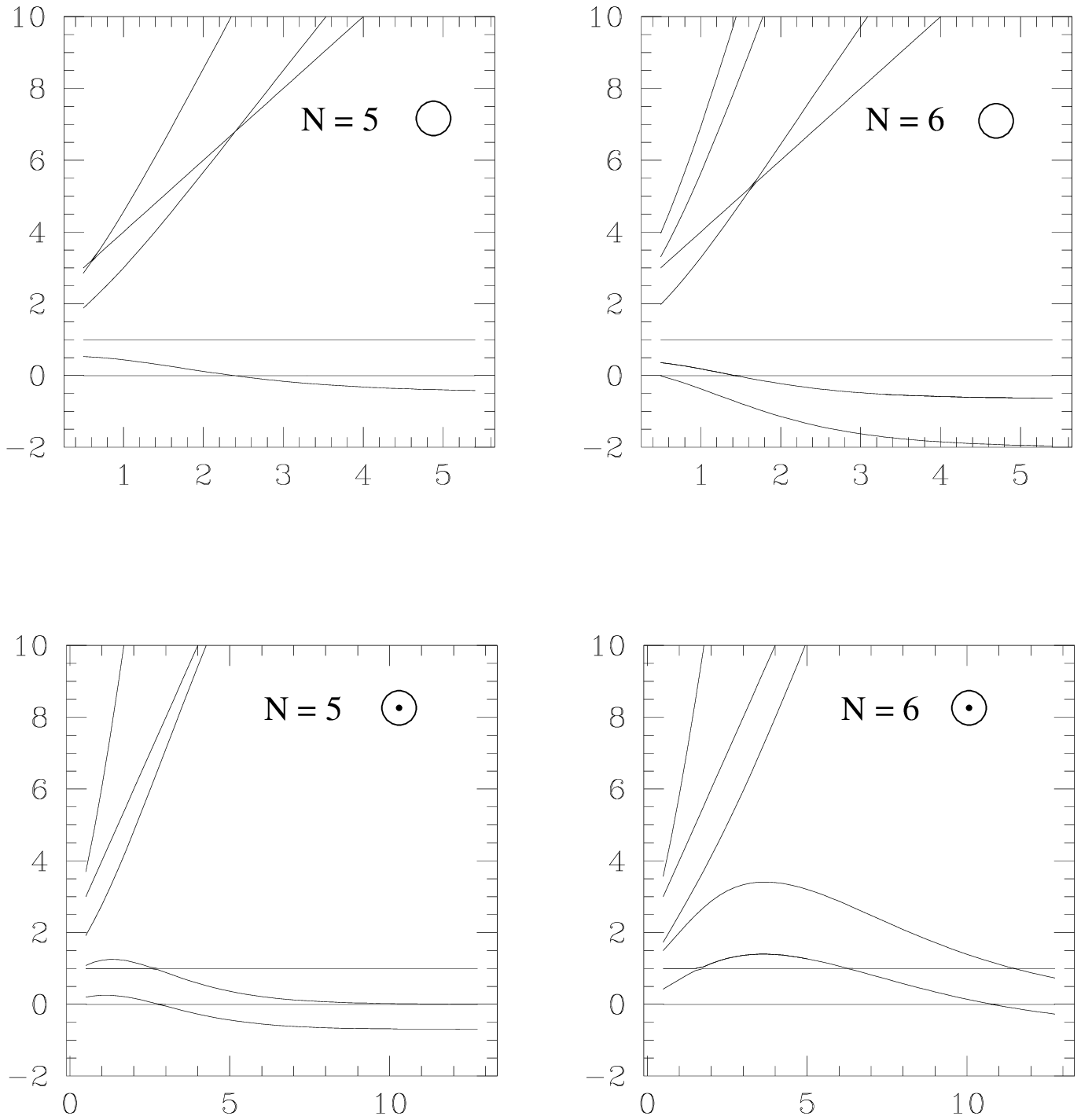}
\begin{tabular}{ll}
{\bf Figure 2:} &
{\sl Eigenvalues of the Hessian of the effective
potential as a function of $\nu$ for}\\
& {\sl $\bigcirc$ and $\bigodot$
configurations and for $N = 5, 6$. Positivity of the eigenvalue}\\
& {\sl indicates a local minimum.}
\end{tabular}
\end{center}

The cases of circle and circle-dot are  particularly simple since
there is only one scale involved, that is, all $s_i^2 = 1, i=2,...,N$ and
$s_1^2 =1$ for the circle and $s_1^2 =0$ for the circle-dot. 

For the circle case, we have, for $N$ particles,
\begin{equation}
\widetilde \phi = \sum_{i=1}^{N-1} s_i^2 +1 = N
\end{equation}
and therefore the energy is given by 
\begin{equation}
E_{\bigcirc}  =\frac{\nu+1}{2\nu}[ 4 g \nu A_{\bigcirc}
N^{\nu+1}]^{\frac{1}{\nu+1}} ,
\end{equation}
where 
\begin{equation}
A_{\bigcirc}(\nu,N) = 
\frac{1}{2^{2\nu+1}} \sum_{k=1}^{N-1}
\frac{1}{\sin^{2\nu}(\frac{\theta_{Nk}}{2})}
=\frac{1}{2^{2\nu+1}} \sum_{k=1}^{N-1}
\frac{1}{\sin^{2\nu}(\frac{k\pi}{N})}. \label{acirc}
\end{equation}
The second equality  follows from the fact that for the symmetric 
configuration on the circle, $\theta_{Nk} = 2\pi (N-k)/N $.

In the case of circle-dot, we have, for $N$ particles ,
\begin{equation}
\widetilde \phi = N-1
\end{equation}
since there are now $N-1$ particles on the circle and one at the centre.
Therefore the energy is given by 
\begin{equation}
E_{\bigodot}  =\frac{\nu+1}{2\nu}[ 4 g \nu A_{\bigodot}
(N-1)^{\nu+1}]^{\frac{1}{\nu+1}} ,
\end{equation}
where 
\begin{equation}
A_{\bigodot}(\nu,N) = A_{\bigcirc}(\nu,N-1)+1. \label{adot}
\end{equation}
The extra 1 on the RHS is the contribution of the particle at the
center.

To ascertain which of the two configurations $\bigcirc$ and
$\bigodot$ has lower energy we look at the following ratio for the 
same number of particles: 
\begin{equation}
f(\nu,N)\equiv 
\left( \frac{E_{\bigcirc}}{E_{\bigodot}} \right)^{\nu+1} 
= \left( \frac{N}{N-1} \right)^{\nu+1}
\frac{\lambda_N^{(\nu)}}{\lambda_{N-1}^{(\nu)}+2^{2\nu+1}} ,
\end{equation}
where
\begin{equation} 
\lambda_N^{(\nu)} \equiv \sum_{k=1}^{N-1} 
\frac{1}{\sin^{2\nu}(\frac{k\pi}{N})}.  \label{lamn}
\end{equation}
Note that in general the ratio $f$
depends only on $N$ and $\nu$ but not on the other parameters of the 
model Hamiltonian. Obviously the circle is a lower 
energy configuration iff $f < 1$. We claim that, for all $~ \nu ~ > ~ 0$, 
\begin{equation}
\begin{array}{lclr}
f(\nu,N) & < & 1  & ~~~ \mbox{for} ~~ N\le 5; \\
f(\nu,N) & > & 1  & ~~~ \mbox{for} ~~ N\ge 6. 
\end{array}
\end{equation}
Further the function $f(\nu,N)$ crosses unity exactly once for $N$
between 5 and 6 and nowhere else. In Fig.3, we show the numerical values of 
the function $f(\nu,N)$ as a function of $N$ for various values of $\nu$. 

\begin{center}
\hspace{0.02cm}
\epsfbox{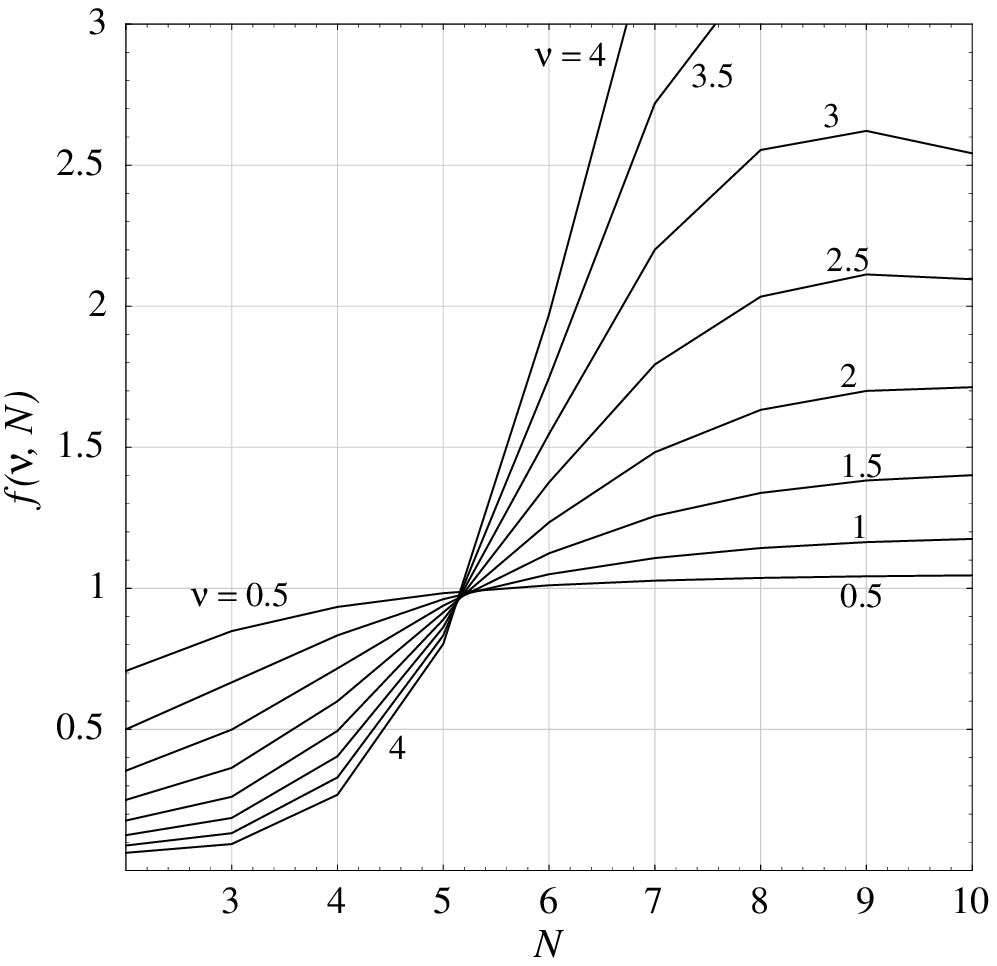}

\mbox{\sl {\bf Figure 3:} The geometric transition:
$f(\nu, N)$ vs $N$ for $\nu = 0.5$ to $\nu = 4$}
\end{center}

The result can be easily checked for
$\nu=1$ since in this case $\lambda_N^{(1)} = (N^2-1)/3$. Therefore,
\begin{equation}
f(1,N)= \frac{N^2(N+1)}{(N-1)(N^2-2N+24)},
\end{equation}
which reproduces the claims made above, for $\nu=1$.
For $\nu<<1 $, again the proof is straightforward and we include it 
here since it will be relevant to the case of logarithmic interaction 
later. For very small $\nu$ we make use of the identity that for any $a$, 
$a^{\nu}\approx 1+\nu \log(a)$.
Using this identity $\lambda_N$ may be written as,
\begin{equation} 
\lambda_N^{(\nu)} \approx \sum_{k=1}^{N-1} (1-2\nu\log(\sin(k\pi/N)) 
=N-1 -2\nu X_N, 
\end{equation}
where 
\begin{equation}
X_N = \log \left[ \prod_{k=1}^{N-1}\sin(\frac{k\pi}{N}) \right] =
  \log \left[ \frac{N}{2^{N-1}} \right].
\end{equation}
Here we have used the identity, 
$\prod_{k=1}^{N-1}\sin(\frac{k\pi}{N}) =\frac{N}{2^{N-1}}$.
Further we also approximate $(N/N-1)^{\nu+1} \approx (N/N-1)(1+\nu 
\log(N/N-1))$.  Substituting for $\lambda_N$ in $f$, we have
\begin{equation}
f(\nu,N) \approx
1+\frac{\nu}{N(N-1)}\mu_N, 
\end{equation}
where
\begin{equation}
\mu_N=[N(N-3)\log(N)-(N-1)(N-2)\log(N-1)] 
\end{equation}
is independent of $\nu$. Clearly, whether $f$ is less than 
or greater than unity depends on whether $\mu_N$ is negative or positive. 
However the properties of $\mu_N$ cannot be dependent on $\nu$. 
It is now easy to see that $\mu_N$ is negative for $N \le 5$ and positive
otherwise. Hence the proof. 

Therefore, for $N\ge 6$, the ground state must have multi-shell 
structure (including possibly $\bigodot$).  Note that the ratio $f>1$ or 
$f<1$ says nothing about whether the $\bigcirc$ or $\bigodot$ is the 
ground state, or even local minimum.  What it does say is that, if $f>1$, 
then $\bigcirc$ {\it can not be the ground state}, independent of its 
stability properties. Therefore, whatever be the ground state  it must 
have at least two shells (with one shell possibly trivial as in 
$\bigodot$). This statement is independent of $\nu$. Thus we conclude 
that the first geometric transition is independent of $\nu, g$ and $ J$ . 
Independent of the  
calculation of $f$, we know from numerical calculations that $\bigcirc$ 
is the ground state for $2\le N\le 5$ and $\bigodot$ is the ground state 
for $6\le N \le 8$ for $\nu < \nu_{max}$ where $\nu_{max}$ is determined 
from the eigenvalues of the Hessian. 

One could have also arrived at the conclusion that for $N=6$ multi-shells 
must form, by studying the eigenvalues of the Hessian. This however is 
harder to compute analytically and also is not conclusive for small 
values of  $\nu$ for which $\bigcirc$ is still a local minimum.  The 
criterion in terms of $f>1$ is, however, easy and conclusive. 
 
It therefore appears that the organisation of many-body clusters in two
dimensions into shells is a robust phenomenon, independent of the nature of
the repulsive two-body interaction and also independent of the Hamiltonian
parameters but dependent only on the number of particles in the cluster. 
In particular, the first geometric transition for the ground state from
circle to circle-dot configuration occurs after $N=5$.  The robustness of
this transition seems to emerge purely from the number theoretic
properties of the ratio of the energies in these two configurations. The
results of this section are, however, restricted to parabolic confinement. 
In the next section, we analyse the effect of confinement on the geometry
of clusters. 

\section{Effect of confinement}

While the parabolic confinement is a simple and popular choice for model 
systems, real systems may be more complicated. In this section, we 
consider a generalisation of the form of the confinement potential and 
study its effects on the cluster properties. For our study we choose a 
form of the confinement potential proposed by Partoens and 
Peeters\cite{partoens} who have studied the influence of the form of the 
confinement potential and the interaction potential using numerical 
methods as well as an approximate Thomson model. To keep the analysis simple 
we do 
not include the  magnetic field. The 
Hamiltonian in  dimensionless units may be written as 
\begin{equation}
H = \sum_{i=1}^{N}\left[\frac{\vec p_i^{~2}}{2} 
+\frac{g'}{2} (r_i^2)^{\gamma}\right] + g\sum_{i,j (j\ne i)} 
\frac{1}{(r_{ij}^2)^{\nu}}, 
\label{ham3}
\end{equation}
where $\gamma$ characterises the confinement potential. While $\gamma=1$ 
corresponds to parabolic potential, we keep it arbitrary. The variational 
equations for extrema of $F$ defined in equation (\ref{eqf}) are 
straightforward to write. As before, one can eliminate the Lagrange 
multiplier in favour of $J$ and $\phi$ (defined in eq.(\ref{phidef})). The 
equilibrium configurations are then determined by,
\begin{eqnarray}
p_{ix} &=& - \frac{J}{\phi} y_i\\
p_{iy} &=&~\frac{J}{\phi}x_i\\  
\left( {g'} \gamma (r_i^2)^{\gamma-1} - \frac{J^2}{\phi^2} \right)\vec r_i &=&  4g\nu  
\sum_{j ( j \ne i)} \frac{\vec r_{ij}}{(r_{ij}^2)^{\nu+1}}. \label{eqrg}
\end{eqnarray}
Taking the cross product in eq.(\ref{eqrg}) with $\vec r_i$ gives $N$ equations which are 
exactly the same as before in Section II. Taking the dot product with 
$\vec r_i$ and dividing by $r_i^2$, however, gives the LHS which is 
{\it dependent} on the index $i$. Thus the parabolic confinement is special in 
this 
regard, since for $\gamma=1$ the LHS is independent of $i$. This fact was 
crucial for the conclusion that the shell structure does not dependent on 
the total angular momentum $J$. In the absence of the magnetic field, we 
may assume that the ground state has $J=0$, in which case the
$i$-dependence of the LHS may be easily eliminated as in Sec.I. The following analysis is 
therefore true for all $J=0$ equilibrium configurations in general and in 
particular for the ground state.

Once again the shell structure is independent of the interaction 
strength. To see this we introduce an auxiliary variable, 
\begin{equation} 
\phi_{\gamma} = \sum_i ({r_i^2})^{\gamma} . 
\label{phidefg} 
\end{equation}
Expressing $ \vec r_i = R \vec s_i$,  $R$ being a
common scale factor (say the radius of the $N^{th}$ particle), we write
\begin{equation}
\phi_{\gamma} = (R^2)^{\gamma} \left[
\sum_{i=1}^{N-1}({s_i^2}^{\gamma}) + 1
\right] 
\equiv  R^{2\gamma} \widetilde \phi_{\gamma} 
\end{equation}
since $s_N^2 = 1$ by choice, as in Sec.II. 
The basic equation determining the configuration space coordinates then 
takes the form
\begin{equation}
g'\gamma \frac{(R^2)^{\nu+\gamma}}{4g\nu} (s_i^2)^{\gamma -1}
\vec s_i =  \sum_{j ( j \ne i)} 
\frac{\vec s_{ij}}{(s_{ij}^2)^{\nu+1}}. \label{eqsig}
\end{equation}
Taking the scalar product with $\vec s_i$ and dividing both sides by
$(s_i^2)^{\gamma} ~(\ne 0)$, we get 
\begin{equation}
g'\gamma\frac{(R^2)^{\nu+\gamma}}{4g\nu} =  \sum_{j ( j \ne i)} \frac{1 -(s_j/ s_i)
\cos(\theta_{ij})}{s_i^{2(\gamma-1)}(s_i^2+s_j^2-2s_is_j\cos(\theta_{ij}))^{\nu+1}}.
\label{eqdotg}  
\end{equation}
Note that the LHS is independent of the particle index $i$ and the RHS is 
different from the parabolic case. However, we still have  $N-1$ 
independent of equations of the type
\begin{eqnarray}
 \sum_{j ( j \ne i)} \frac{1 -(s_j/ s_i)
\cos(\theta_{ij})} 
{s_i^{2(\gamma-1)}(s_i^2+s_j^2-2s_is_j\cos(\theta_{ij}))^{\nu+1}}    
~~~~~~~~~~~~~~~~~~~~~~~~~~~~~~~~~~~~\nonumber \\
=\sum_{j\ne k} \frac{1 -(s_j/ s_k)
\cos(\theta_{kj})}
{s_k^{2(\gamma-1)}(s_k^2+s_j^2-2s_ks_j\cos(\theta_{kj}))^{\nu+1}},
~~\forall ~~~k\ne i. \label{eqdtg}
\end{eqnarray}
Further by taking the cross product with $\vec s_i$ and dividing by
$s_i$, we get, 
\begin{equation}
\sum_{j ( j \ne i)} \frac{s_j
\sin(\theta_{ij})}{(s_i^2+s_j^2-2s_is_j\cos(\theta_{ij}))^{\nu+1}} = 0,
\label{eqcrsg}  
\end{equation}
which is independent of $\gamma$ and therefore is identical to the 
parabolic case. This equation provides a further set of $N$ conditions on 
the internal
coordinates $\vec s_i$. 
Together
eqs.(\ref{eqdtg},\ref{eqcrsg}) provide the $2N-1$ necessary
equations for determining the $s_i$ and the angles $\theta_i$ 
and are completely independent of  $g$ and $g'$.

These, the strengths of confinement and interaction potentials, determine 
the overall scale $R$ through
eq.(\ref{eqdotg}).  Setting $i=N$ in eq.(\ref{eqdotg}), we get
\begin{equation}
g'\gamma\frac{(R^2)^{\nu+\gamma}}{4g\nu}=\sum_{j=1}^{N-1} \frac{1
-s_j\cos(\theta_{Nj})}{(1+s_j^2-2s_j\cos(\theta_{Nj}))^{\nu+1}} 
\label{eqscaleg}  
\end{equation}
Thus we have the
result that the {\it geometry } 
or the shell structure of the equilibrium configuration is {\it independent} of
the parameters of the Hamiltonian, which only restrict the overall size
of the system.

Next we compute the energy of the equilibrium configurations. To this end 
we note that the auxiliary variable $\phi_{\gamma}$ defined in
eq.(\ref{phidefg}) is related to the two-body potential energy by,
\begin{equation}
g'\gamma\phi_{\gamma}=  2g\nu  \sum_{i,j ( j \ne i)} \frac{1}
{(r_{ij}^2)^{\nu}}, \label{eqphig}
\end{equation}
where the RHS is proportional to the potential energy due to
interaction. Therefore, 
\begin{equation}
\gamma[g'\frac{1}{2}\sum _i (r_i^2)^{\gamma}] =\nu [ g \sum_{i,j ( j \ne i)} 
\frac{1} {(r_{ij}^2)^{\nu}}],
\end{equation}
which implies that the $\gamma$ times the confinement energy is $\nu$ times 
the interaction 
energy. This is the generalised ``virial theorem" valid for arbitrary 
confinement. We have, for the energy at the extrema,
\begin{equation}
E =\frac{\nu+\gamma}{2\nu}g'\phi_{\gamma}.
\label{eqeng}
\end{equation}
This expression for 
energy is valid for all equilibrium configurations. The scale $R$ 
may be calculated from eq.(\ref{eqscaleg}) and is given by
\begin{equation}
R^2 = [\frac{ 4 g \nu}{g'\gamma} A(\nu,N)]^{\frac{1}{\nu+\gamma}} ,
\end{equation}
where
\begin{equation}
A(\nu,N) \equiv \sum_{j=1}^{N-1} \frac{1
-s_j\cos(\theta_{Nj})}{(1+s_j^2-2s_j\cos(\theta_{Nj}))^{\nu+1}}. 
\end{equation}
and is independent of $\gamma$. This definition of $A(\nu,N)$ is the same 
as in the parabolic case.

We now consider the effect of confinement on the geometric transition for 
$N$ from 5 to 6. As shown in the previous section for parabolic 
confinement,  the configuration changes from  circle to circle-dot 
independent of $\nu$. We first give expressions for the energy for these 
two configurations for any $N$. 

For the circle case, we have, for $N$ particles,
\begin{equation}
\widetilde \phi_{\gamma} = [\sum_{i=1}^{N-1} s_i^{2\gamma} +1] = N
\end{equation}
and therefore the energy is given by 
\begin{equation}
E_{\bigcirc}  =\frac{\nu+\gamma}{2\nu} g' \left[ \frac{4 g \nu}{g'\gamma} 
A_{\bigcirc} N^{\frac{\nu+\gamma}{\gamma}} \right]^{\frac{\gamma}{\nu+\gamma}} ,
\end{equation}
where $A_{\bigcirc}(\nu,N)$ is given in eq.(\ref{acirc}).

In the case of circle-dot, we have, for $N$ particles,
\begin{equation}
\widetilde \phi_{\gamma} = N-1
\end{equation}
since there are now $N-1$ particles on the circle and one at the centre.
Therefore the energy is given by 
\begin{equation}
E_{\bigodot}  =\frac{\nu+\gamma}{2\nu}g' \left[ \frac{4 g \nu}{g'\gamma} 
A_{\bigodot} (N-1)^{\frac{\nu+\gamma}{\gamma}} \right]^{\frac{\gamma}{\nu+\gamma}} ,
\end{equation}
where 
\begin{equation}
A_{\bigodot}(\nu,N) = A_{\bigcirc}(\nu,N-1)+1 
\end{equation}
as before. 

To ascertain which of these two configurations $\bigcirc$ and
$\bigodot$ has lower energy we look at the ratio for the 
same number of particles, $N$, 
\begin{eqnarray} 
f(\gamma, \nu,N) \equiv 
\left( \frac{E_{\bigcirc}}{E_{\bigodot}} \right)^{\frac{\nu+\gamma}{\gamma}} 
& = & 
\left( \frac{N}{N-1} \right)^{\frac{\nu+\gamma}{\gamma}}
\frac{\lambda_N^{(\nu)}}{\lambda_{N-1}^{(\nu)}+2^{2\nu+1}} \nonumber
\\
& = &  {\left( \frac{N}{N - 1} \right)}^{\frac{ \nu (1 - \gamma)}{\gamma}}
  f(1,\nu,N) 
\end{eqnarray}
where $\lambda_N^{(\nu)}$ is given in eq.(\ref{lamn}) in the previous 
section. The effect of the confinement is therefore entirely contained in 
the prefactor whose exponent contains $\gamma$.  
\begin{center}
\epsfbox{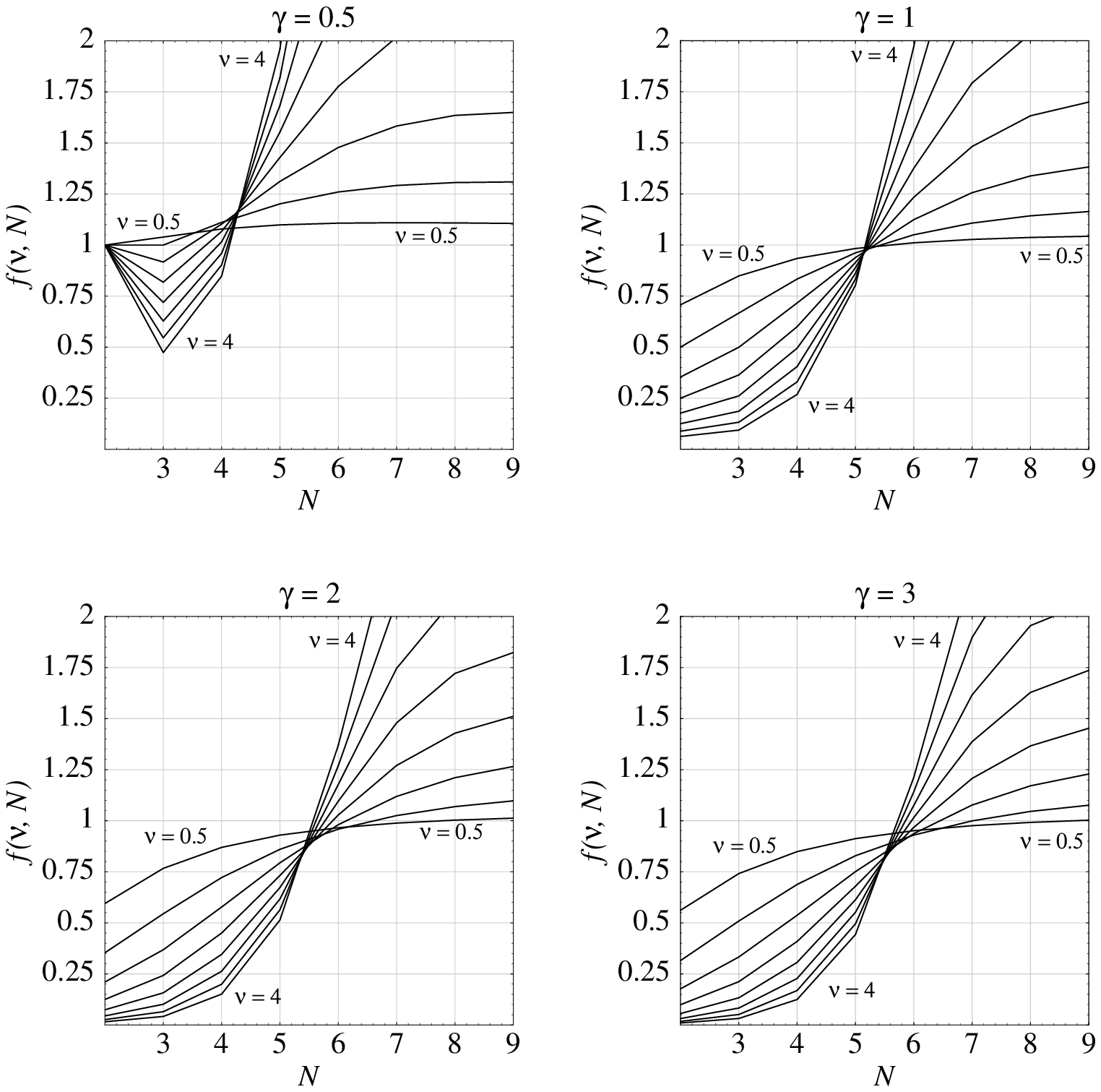}
\begin{tabular}{ll}
{\bf Figure 4:} &
{\sl Effect of Confinement: $ f(\gamma, \nu, N)$ vs
$N$ for different values
of $\gamma$.} \\
&{\sl The values  of $\nu$ range from $0.5$ to $4$ as in Fig.
3.}
\end{tabular}
\end{center}

In Fig.4, we show the 
behaviour of $f$ as a function of $N$ for typical values of $\nu$ and 
$\gamma$. The pattern seen in Fig.4 may be summarised in two parts:
\begin{enumerate}
\item{ 
For $\gamma < 1$, the first geometric transition occurs before $N=5$. As 
$\nu$ increases the transition moves towards $N=5$ but never beyond. In 
fact there exists $\nu_{min}$ after which the transition always occurs 
between 4 and 5 for $N$. }
\item{ For $\gamma > 1$, the first geometric transition occurs  after $N=6$. As 
$\nu$ increases the transition moves towards $N=5$. For $\nu > \nu_{min}$ 
the transition always occurs between 5 and 6 for $N$.}
\end{enumerate}
 In the case of 
Coulomb interaction it was already known that the transition occurs 
either before $N=5$ ($\gamma<1$)  or after $N=6$ 
($\gamma>1$)\cite{partoens}. However 
the fact that the transition always occurs around $N=5$ after some 
$\nu_{min}$  is, to our knowledge, a new result.  

\section{Effect of three-body perturbations}

In this section we consider a model Hamiltonian with a three-body 
interaction. While 
classically three-body interactions are not naturally introduced, quantum 
correlations can of course introduce effective many-body terms.  
The primary motivation for considering the three-body term here 
is to see if the shell structure survives under this perturbation. 
The  effect of the three-body term is interesting and 
non-trivial on the first geometric transitions. The model Hamiltonian, 
in dimensionless units,  is given by
\begin{equation}
H = \sum_{i=1}^{N}\left[\frac{\vec p_i^{~2}}{2} 
+\frac{1}{2} r_i^2\right] + \frac{g_1}{2}\sum_{i,j (j\ne i)} 
\frac{1}{r_{ij}^2} + \frac{g_2}{2} \sum_{i,j,k (i\ne j\ne k)} 
\frac{\vec r_{ij}.\vec r_{ik}}{r_{ij}^2 r_{ik}^2}, \label{ham4}
\end{equation}
where $g_1$ and $g_2$ are in general kept arbitrary. Notice that when 
$g_2=0$, this reduces to the model Hamiltonian analysed in section II 
with $\nu=1$. The quantum mechanics with this model Hamiltonian ($g_2=0$) 
has some 
interesting applications to quantum dots and is considered in detail in 
ref.\cite{johnson}. The model Hamiltonian given above has a very 
interesting 
limit. When $g_1=g_2=g^2$, the quantum mechanical ground state and 
an infinite tower of states may be solved for exactly \cite{date,khare}. 
Solving  the 
Schroedinger equation, $H\psi_0=E_0\psi_0$, for the ground state we obtain
\begin{equation}
\psi_0 = \prod_{i<j} |\vec r_{ij}|^g \exp(-{1\over 2}\sum_i r_i^2),
\end{equation}
and the ground state energy is given by
\begin{equation}
E_0 = N +{1\over 2} g N(N-1).
\end{equation}
In what follows we keep $g_1$ and $g_2$ arbitrary and comment on the case 
$g_1=g_2$ later.  

The classical analysis of the model Hamiltonian is done as before by 
extremising the Hamiltonian in the full phase space. As in the previous 
section we concentrate  only on the case where the total angular momentum,
 $J=0$ , which 
implies that  $\vec p_i =0$ and 
the equilibrium configurations are obtained as solutions of the 
equations 
\begin{equation}
\vec r_i = 2g_1 \sum_{j ( j \ne i))} \frac{\vec r_{ij}}{r_{ij}^4} 
-g_2\sum_{j,k(j\ne k\ne i)} \left[ \frac{\vec r_{jk}}{r_{jk}^2 r_{ik}^2} + 
\frac{\vec r_{ik}}{r_{ik}^2 r_{ij}^2}\right]
+2g_2\sum_{j,k(j\ne k\ne i)} \left[ \frac{\vec r_{jk}.\vec r_{ik}}{r_{jk}^2 
r_{ik}^4}\vec r_{ik} + \frac{\vec r_{ik}.\vec r_{ij}}{r_{ik}^2 
r_{ij}^4} \vec r_{ij}\right].
\end{equation}
 Again these equilibrium configurations may be local 
minima/maxima or saddle points. 

The general analysis of these equations proceeds as in section II.  Since
there are two coupling constants $g_1$ and $g_2$, the shell structure does
depend on these parameters, unlike in the earlier models, except in
the special case when the
two are equal. Taking the dot product on both 
sides of the above equation with $\vec r_i$ and summing over the 
index $i$, we get,
\begin{equation}
\sum_{i=1}^{N}\frac{1}{2} r_i^2 = \frac{g_1}{2}\sum_{i,j ( j \ne i)} 
\frac{1}{r_{ij}^2} + \frac{g_2}{2} \sum_{i\ne j\ne k} 
\frac{\vec r_{ij}.\vec r_{ik}}{r_{ij}^2 r_{ik}^2}.
\end{equation}
Therefore the confinement energy is again equal to the interaction energy 
including the two- and three-body terms. Using this, the energy in any 
equilibrium configuration may be calculated:
\begin{equation}
E = \phi = R^2\sum_{i=1}^N s_i^2 = R^2 \widetilde \phi,
\end{equation}
where $s_i$ are internal variables and $R$ is the overall scale, which 
may be taken to be the distance of the farthest  particle from 
the  origin. 
We now specialise to specific configurations namely the circle and the 
circle-dot. It may be easily checked that these configurations are 
allowed equilibrium configurations for all $N$. These, however,  need not 
be  local minima for all $N$. 
For these two cases, only the overall scale factor $R$ is to
be determined. The angles $\theta_{ij}/2$, as before, are simply 
multiples of 
$\pi/N$ and $\pi/(N -1)$ respectively. 

For the circle case, we have, for $N$ particles,
\begin{equation}
\widetilde \phi = [\sum_{i=1}^{N-1} s_i^2 +1] = N
\end{equation}
and therefore the energy, after some algebra, is given by 
\begin{equation}
E_{\bigcirc}  = N R^2 = 
N \sqrt{\frac{(N-1)}{12} [ g_1(N+1) + 2g_2(N-2)]}
\label{ecirc}
\end{equation}

In the case of circle-dot, we have, for $N$ particles, 
\begin{equation}
\widetilde \phi = N-1
\end{equation}
since there are now $N-1$ particles on the circle and one at the centre.
Therefore the energy is given by 
\begin{equation}
E_{\bigodot}  =(N-1) R^2 = (N-1)\sqrt{ g_1(\frac{N(N-2)}{12}+2) + g_2 
(\frac{(N-2)(N-3)}{6}+(N-3))}.
\label{ecdot}
\end{equation}

Interestingly, in the limit $g_1=g_2=g^2$, we have the important result
\begin{equation}
E_{\bigcirc}=E_{\bigodot} = {1\over 2}g N (N-1),
\end{equation}
which is the same as the quantum mechanical ground state energy without 
the zero point fluctuation. The two configurations are also degenerate in 
this limit.  We may therefore use either of these two classical 
configurations as a starting point for calculating the quantum 
corrections. Since the quadratic fluctuations reproduce the zero point 
energy the higher order corrections must be either vanishing or 
spurious. 

Coming back to the general case, $g_1 \ne g_2$, we would like to examine 
the effect of three-body terms in the interaction on the first geometric 
transition from circle to circle-dot configuration. Note that in the 
absence of the three-body terms, circle is the ground state up to 5 
particles and circle dot is the ground state for $N=6$. 
To ascertain which of these two configurations $\bigcirc$ and
$\bigodot$ has lower energy in the presence of the three-body perturbations,  
it is sufficient to look at the ratio of the energies for the same number of 
particles:
 \begin{equation}
f(N) \equiv
\left(\frac{E_{\bigcirc}}{E_{\bigodot}} \right)^{2} 
 = \left(\frac{N}{N-1} \right)^{2}
\frac{ N^2 [N+ 1 + 2b( N - 2) ]}
     {(N -1)[ N(N-2)+24 +2b(N-3)(N+4)]}. \label{f3b}
\end{equation}
 where $b$ is the ratio $g_2/g_1$ of the strengths of the two terms.
Obviously the circle is a lower 
energy configuration iff $f < 1$. 

\vskip 1cm
\begin{center}
\epsfbox{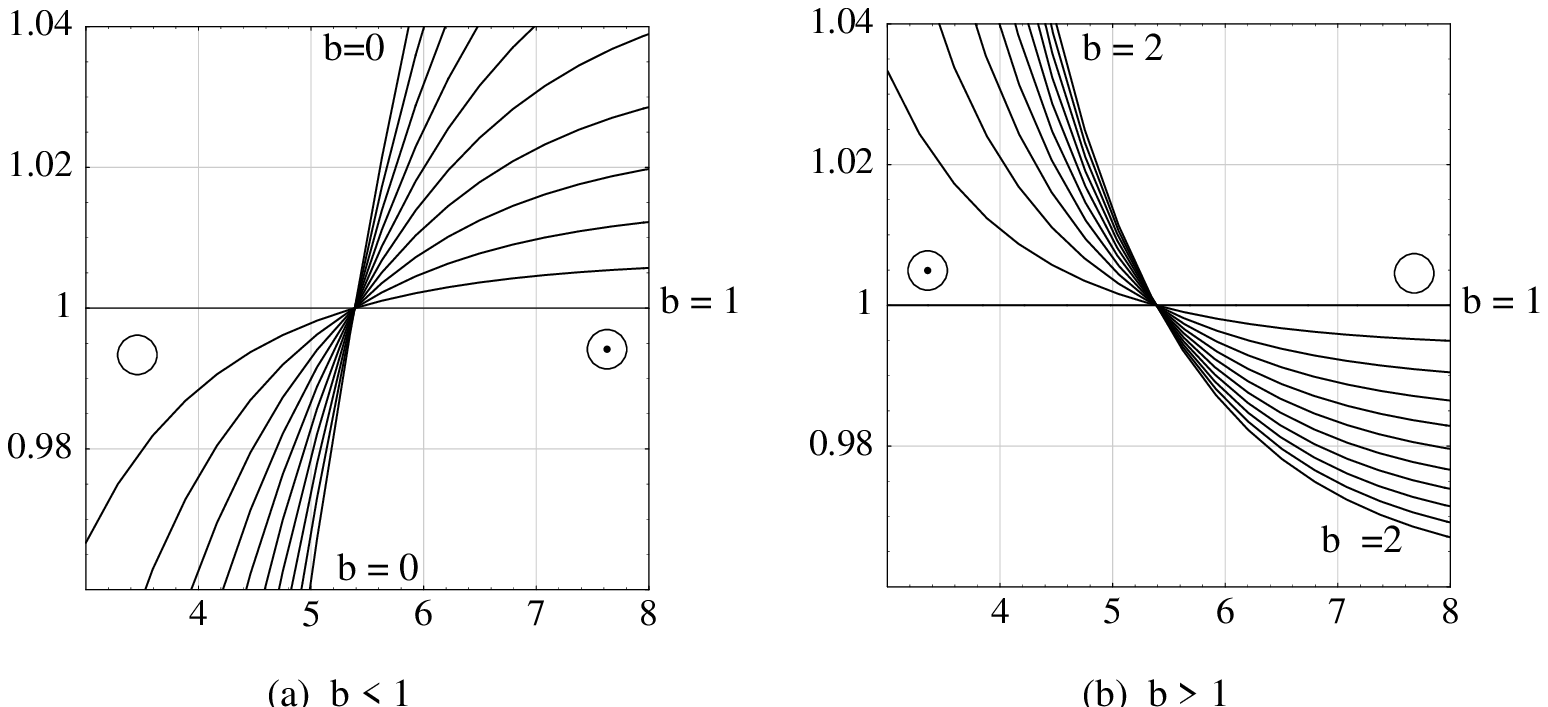}
\begin{tabular}{ll}
{\bf  Figure 5:}&
{\sl $f(N)$ vs $N$ for various values of
the ratio of the three-body to two-body}\\
& {\sl strength, $b = g_2/g_1$.
In (a) $b$ varies from $0$ to $1$ in steps of $0.1$ and in }\\
& {\sl (b) $b$ varies
from $1$ to $2$ in steps of $0.1$.}
\end{tabular}
\end{center}

In Fig. 5, we show the numerical values of 
$f(N)$ as a function of $N$ for various values of  
$ b = g_2/g_1$. 
Some interesting facts emerge from Fig. 5. First, at $b=1$, the ratio 
is 1 since the two configurations are degenerate in this limit. 
For $b<1$, circle has lower energy for $N\le 5$ and therefore $f<1$. 
The ratio crosses unity when $N$ changes from 5 to 6 at exactly the same 
point  no matter what the value of $b$ is. It is as if the three-body 
perturbation has no effect on this geometric transition. The energies 
of course explicitly depend on the coupling constants. However,  
for $b >1$, circle has higher  energy for  $N\le 5$ and therefore 
$f>1$. Again the ratio crosses unity between 5 and 6 and for all $N>6$, 
 circle configuration has lower energy. 
This can be seen analytically also, from eq.(\ref{f3b}).

However, it turns out neither the circle nor the circle-dot configuration
can be the ground state when $b>1$, that is when the three-body term
dominates the two-body term in the Hamiltonian. In fact the ground state is
one when all particles on a line, where the particle positions are determined
by the zeros of the Hermite polynomial of order $N$\cite{dgm}. We must,
however, point out that the above results are specific to the form
of the Hamiltonian. Whether three-body interactions, in general, have
a similar effect is a difficult question to answer.

\section{Discussion}

To summarise, we have discussed the shell structure of particle clusters
in the presence of external confinement. To begin with, we have assumed
harmonic confinement and the interaction is two-body. Following the
analysis in section II, it appears that the organisation of many-body
clusters in two dimensions into shells is a robust phenomenon, independent
of the nature of the repulsive two-body interaction and also independent
of the Hamiltonian parameters but dependent only on the number of
particles in the cluster.  In particular, the first geometric transition
for the ground state from circle to circle-dot configuration occurs after
$N=5$.  The robustness of this transition seems to emerge purely from the
number theoretic properties of the ratios of the energies in these two
configurations.  A different type of confinement may in fact destroy the
robustness of this transition for small $\nu$. The interesting point,
however, is that after some critical $\nu_{max}$, once again the results look
similar to that of the parabolic confinement.

Further, the presence of three-body terms in the
interaction does not seem to affect the geometric transition as long as it
remains a weak perturbation. Interestingly enough, within the model
Hamiltonian we have analysed, there exists a critical point when the
geometric transition changes dramatically as the perturbation strength is
varied. Most importantly, the classical energy is the same as the quantum 
mechanical energy (without the zero point fluctuations) in the limit when 
the strengths of the two- and three-body terms are equal.  It would be 
interesting to check if this is true of other many-body models where the 
quantum ground state is exactly known.  

While the existence of the shell structure is numerically well 
established, the analytical understanding of the existence of the shells 
is lacking. It would be interesting to know how many distinct extremal 
configurations there are . While the first transition seems to indicate 
the formation of shells, a deeper understanding of the Mendeleev table is 
clearly called for. It would also be interesting to study quantum 
corrections using these classical configurations as a starting point. 
Such a study would be of relevance to systems in which quantum
corrections are non-negligible, as in, say, quantum dot systems.
Some of the afore-mentioned questions are being probed further.

\acknowledgments

We thank R. Balasubramanian and D. Surya Ramana for help with number
theoretic identities. We thank P.P. Divakaran for pointing out the 
reference to Myer's work. We also thank A.P. Balachandran, Avinash 
Khare, Madan Rao and Surajit Sengupta for many helpful discussions. 

\newpage

\appendix{\bf {Appendix A: Logarithmic Interactions}}
\setcounter{equation}{0}
\renewcommand{\theequation}{A.\arabic{equation}}

We briefly discuss the case of logarithmic interaction potential with 
harmonic confinement in this appendix. In some quasi-two-dimensional 
systems like quantum dots, this may be closer to the physical situation 
than the power-law interactions.  Specifically 
the Hamiltonian, in scaled units, may be written as 
\begin{equation}
\frac{H}{\hbar\omega} = \sum_{i=1}^{N}\left[\frac{\vec p_i^{~2}}{2} 
+ \frac{\vec r_i^2}{2}\right] - g\sum_{i,j ( j \ne i)} 
\log \left( {\frac{r_{ij}^2}{\rho^2}} \right), 
\label{hamlog}
\end{equation}
where $\rho$ is an arbitrary length parameter chosen such that $\rho^2 > 
r_{ij}^2 ~~\forall~~i, j$. This is to ensure that the interaction potential is repulsive 
for all distances in the cluster. For simplicity, we do not include 
magnetic field here, though the analysis of Sec. I goes through 
identically in this case also. The conditions for the equilibrium are 
given by
\begin{equation}
\vec r_i =  4g  \sum_{j ( j \ne i)} \frac{\vec 
r_{ij}}{r_{ij}^2}. 
\end{equation}
Again, by setting 
$\phi = R^2[\sum_{i=1}^{N-1} s_i^2 +1] \equiv R^2 \widetilde \phi$,
it is easy to demonstrate that the shell structure is independent of $g$ 
(and also of the magnetic field, when included) as in Sec. I of the paper. 
Note that the equation above is 
manifestly independent of the arbitrary scale $\rho$ introduced in the 
Hamiltonian. Further simplification occurs if we note that for any 
equilibrium configuration the auxiliary variable, $\phi$ is given by
\begin{equation}
\phi=  2g N(N-1)
\end{equation}
and the energy for the equilibrium configuration is given by
\begin{equation}
E= gN(N-1)[ 1+ \log(\rho^2)]-g\sum_{i,j ( j \ne i)} 
\log(r_{ij}^2).
\end{equation}

We now calculate the energy for specific configurations, namely circle 
and the circle-dot. For the circle case, we have, for $N$ particles 
distributed symmetrically,
\begin{equation}
E_{\bigcirc}  = gN(N-1)[ 1+ \log(\rho^2)-\log(2gN(N-1))]
+gN(N-3)\log(N).
\end{equation}

In the case of circle-dot, for $N$ particles 
with $N-1$ particles distributed symmetrically on the circle and one at the 
centre, the energy is given by 
\begin{equation}
E_{\bigodot}  =
gN(N-1)[ 1+ \log(\rho^2)-\log(2gN(N-1))]
+g(N-1)(N-2)\log(N-1).
\end{equation}

It is not possible in general to calculate these energies unless we 
specify the arbitrary scale $\rho$. However, to ascertain which of these 
two  configurations $\bigcirc$ and
$\bigodot$ has lower energy it is sufficient to look at the difference for 
the same number of particles, that is,
\begin{equation}
\Delta E = E_{\bigcirc} -E_{\bigodot} = g[N(N-3)\log(N) - 
(N-1)(N-2)\log(N-1)] = g\mu_N.
 \end{equation}
where $\mu_N$ is already defined in Sec. II. 
Clearly the circle configuration has lower (higher) energy if $\mu_N$ is 
negative (positive). Note that this statement is independent of the 
arbitrary scale and also the interaction strength.  
It is now easy to see that $\mu_N$ is negative for $N \le 5$ and positive
otherwise. Therefore the first geometric transition from circle to 
circle-dot occurs when the number of particles changes from 5 to 6, as in 
the case of power-law potentials.  

\newpage
\appendix{\bf {Appendix B: Numerical Simulations}}

\setcounter{equation}{0}
\renewcommand{\theequation}{B.\arabic{equation}}

In this appendix we briefly discuss the numerical simulations carried out 
for multi-shell configurations. The analysis is carried out for the 
Hamiltonian given in Section II for parabolic confinement and may be 
adapted easily for other model systems discussed in Sections III and IV. 
Also collected are some results on the eigenvalues of the Hessian of
the effective potential at the 
extrema. From eq.(\ref{eqr}) it is easy to see that the same may be 
obtained as $\vec \nabla V_{eff}(\vec r_i)=0$, where
\begin{equation}
V_{eff}= \frac{1}{2}\left[ (1+\alpha^2)\phi +\frac{J^2}{\phi} \right]
+ g \sum_{i,j (j\ne i))} \frac{1}{r_{ij}^2}.
\label{Veff}
\end{equation}
The equilibrium configurations thus are obtained by {\it minimising} 
$V_{eff}$ numerically.  For this we use the version of the conjugate gradient 
method\cite{numrec} which uses line minimisation. Since $V_{eff}$ is 
positive definite and we use line minimisation, we are guaranteed to reach 
the extrema which are either local minima or some times saddle points but 
never local maxima. 

For a given number of particles, we choose several initial configurations 
graphically and compare the energies at the local minima.  We also check 
that a local extremum reached is actually a local minimum by computing 
the eigenvalues of the Hessian of $V_{eff}$ at the extremum.  We verify the 
``Mendeleev" table of Bedanov and Peeters\cite{bedanov} completely for the 
special case of  Coulomb interaction, i.e. when $\nu=1/2$. 

The Hessian of $V_{eff}$ can be computed easily. It turns out that of the 
$2N$ eigenvalues (for $N$ particles), four can be obtained exactly. These 
also provide a check on the numerically determined eigenvalues. That the 
numerically determined eigenvalues contain the exact eigenvalues is also 
verified. Since the geometry of the extrema is independent of $J$, this 
analysis is done for zero magnetic field and $J=0$.  At an arbitrary 
point in the configuration the elements of the  Hessian matrix of $V_{eff}$
 are given by
\begin{eqnarray}
\frac{\partial^2V_{eff}}{\partial x_i\partial x_j} = 
(1+\alpha^2)\delta_{ij} -4g\nu \left[\delta_{ij} \sum_{k (k\ne i)} 
\frac{y^2_{ik} -(1+2\nu)x^2_{ik}}{(r^2_{ik})^{\nu+2}})  
-(1-\delta_{ij})\frac{y^2_{ij} -(1+2\nu)x^2_{ij}}{(r^2_{ij})^{\nu+2}}
\right] \\
\frac{\partial^2V_{eff}}{\partial y_i\partial y_j} = 
(1+\alpha^2)\delta_{ij} -4g\nu \left[\delta_{ij} \sum_{k (k\ne i)} 
\frac{x^2_{ik} -(1+2\nu)y^2_{ik}}{(r^2_{ik})^{\nu+2}})  
-(1-\delta_{ij})\frac{x^2_{ij} -(1+2\nu)y^2_{ij}}{(r^2_{ij})^{\nu+2}}
\right] 
\end{eqnarray}
\begin{equation}
\frac{\partial^2V_{eff}}{\partial x_i\partial y_j} = \nonumber 
\frac{\partial^2V_{eff}}{\partial y_i\partial x_j} = 
8g\nu (\nu + 1) \left[\delta_{ij} \sum_{k (k\ne i)} 
\frac{y_{ik}x_{ik}}{(r^2_{ik})^{\nu+2}}  
-(1-\delta_{ij})\frac{y_{ij}x_{ij}}{(r^2_{ij})^{\nu+2}} \right]
\end{equation}
The eigenvalue equation for this matrix is
\begin{eqnarray}
 \left( \begin{array}{cc}
       \frac{\partial^2V_{eff}}{\partial x_i\partial x_j} &
       \frac{\partial^2V_{eff}}{\partial x_i\partial y_j} \\
       \frac{\partial^2V_{eff}}{\partial y_i\partial x_j} &
       \frac{\partial^2V_{eff}}{\partial y_i\partial y_j}
       \end{array} \right)
\left( \begin{array}{c}
       X_j \\ Y_j 
       \end{array} \right)  =
 ( 1+ \alpha^2)\mu \left( \begin{array}{c} 
       X_i \\ Y_i  
       \end{array} \right)
\end{eqnarray}
By grouping $(X_i, Y_i)$ as two-dimensional vectors $\vec R_i$, one can 
write the matrix equation conveniently as 
\begin{equation}
\left( \frac{1 + \alpha^2}{4 g \nu} \right)
    (1 - \mu) \vec R_i =
\sum_{j ( j \ne i)} \frac{\vec R_{ij}}{(r_{ij}^2)^{\nu + 1}}
- 2 (1+\nu) \sum_{j ( j \ne i)} 
    \frac{ \vec r_{ij}( \vec R_{ij}.\vec r_{ij})}
         {(r_{ij}^2)^{\nu + 2}}
~~~ \forall ~~~ i = 1 ... N.
\end{equation}
At the  extrema, $\vec r_i$ are of course solutions of the equations
\begin{equation}
\left( \frac{1 + \alpha^2}{4 g \nu} \right)
   \vec r_i =
\sum_{j ( j \ne i)} \frac{\vec r_{ij}}{(r_{ij}^2)^{\nu + 1}}
~~~~ \forall~~~ i = 1 ... N.
\end{equation}
From these equations, four exact eigenvalues and eigenvectors can be 
deduced immediately:
\begin{enumerate}
\item{ Both the equilibrium equations and the eigenvalue equations are 
manifestly rotationally covariant.  One therefore expects $\mu=0$ to be 
an eigenvalue with the corresponding eigenvector $\vec R_i$ denoting the 
rotation of the $\vec r_i$.  This is indeed the case. Explicitly,
\( \mu = 0 \) and \( \vec R_i = \rho \hat k \times \vec r_i ~~\forall~~i =
1 ... N \)
solves the eigenvalue equation when $\vec r_i$ satisfy the equilibrium 
equations.  Here $\rho$ is an arbitrary, non-zero factor since the 
eigenvalue equations are homogeneous. 
}
\item{ If $\vec R_i = \rho \vec r_i$ for all $i=1...N$, then, using 
the equilibrium equations, it follows that the eigenvalue equations are 
satisfied with $\mu =2(1+\nu)$.  Thus the equilibrium configurations 
themselves constitute an eigenvector with an eigenvalue which depends 
only $\nu$. In particular it is common to {\it all extrema for every $N$}. 
}
\item{ It is immediately obvious that if $\vec R_i = \vec a~~\forall~~i$, then 
the RHS of the eigenvalue equations vanishes and $\mu=1$ is the 
corresponding eigenvalue.  As there are two independent choices for the 
two dimensional vector $\vec a$, this eigenvalue is doubly degenerate.  
Note that this eigenvalue is totally independent even of the equilibrium 
equations. 
}
\end{enumerate}

As all these eigenvalues are common to all extrema, the remaining 
eigenvalues must distinguish saddle points, local minima and local 
maxima. Apart from providing a check on numerical simulations, these should 
also prove useful in computing the semi-classical corrections
to the classical energies.

\end{document}